\documentclass[11pt]{article}
\usepackage{amssymb,amsmath,amsthm,hyperref}
\usepackage{algorithm}
\usepackage{algpseudocode}

\hypersetup{
    colorlinks,%
    citecolor=blue,%
    filecolor=blue,%
    linkcolor=blue,%
    urlcolor=blue
}

\newcommand{\lrf}{\mathcal{L}_{\mathbb{R}_{\mathcal{F}}}}
\usepackage{fullpage}
\theoremstyle{definition}
\newtheorem{definition}{Definition}[section]
\newtheorem{example}{Example}[section]
\newtheorem{theorem}{Theorem}[section]

\newtheorem{corollary}{Corollary}[section]

\newtheorem{goal}{Goal}

\newcommand{\N}{\mathbb{N}}
\newcommand{\dom}{\mathrm{dom}}
\newcommand{\R}{\mathbb{R}}

\title{\bf Descriptive Control Theory: A Proposal}
\author{Sicun Gao}
\date{}

\begin{document}
\maketitle
\thispagestyle{empty}
\begin{abstract}
Logic is playing an increasingly important role in the engineering of real-time, hybrid, and cyber-physical systems, but mostly in the form of posterior verification and high-level analysis. The core methodology in the design of real-world systems consists mainly of control theory and numerical analysis, and has remained mostly free of logic and formal approaches. As a result, besides facing extreme difficulty in guaranteeing the reliability of these systems, engineers are also missing out on the computational power of logic-based methods that has greatly advanced in the past decades. To change this situation, we need a logical and computational foundation for control theory. The name ``descriptive control theory" emphasizes the overarching theme of using logic to express, analyze, and solve problems in control theory. If the program is successfully carried out, logical approaches will significantly extend existing engineering methods towards a unified methodology for handling nonlinear and hybrid systems, and bring design automation and reliability to an unprecedented level in the broad field of engineering. 
\end{abstract}

\vspace{.6cm}

\section{Introduction}

To engineer a system is, ideally, to prove a theorem. While such a statement can often be made exact for software systems in the sense of Curry-Howard, it seems hardly applicable to the engineering of {\em real-world systems}: systems that physically engage us in safety-critical ways, from airplanes to nuclear plants to cardiac pacemakers. Admittedly, the study of formal methods for real-time, hybrid, and cyber-physical systems has given logic an increasingly important role in real-world system engineering, but mostly in the form of posterior verification or very high-level analysis. The core methodology in the design of these systems consists mainly of control theory and numerical analysis, and has remained mostly free of logic and formal approaches. An obvious consequence is the well-known difficulty in guaranteeing the reliability of these systems. An equally fundamental problem, which appears less often recognized, is that engineers are missing out on the computational power of logic-based methods that has greatly advanced in the past decades. I believe that logic and formal approaches, backed by their underlying computational engines, have the power of bringing design automation and reliability to an unprecedented level in the broad field of engineering. Ultimately, the process of engineering should automatically produce logical derivations of correctness of the constructed systems, and in this process, logical decision procedures should play a part as indispensable as calculus, linear algebra, and optimization. When such a unification of methodologies happens, the systems that we are able to build will be orders of magnitude more complex and reliable, and the physical world and the computational world will quickly converge. 

Towards such a goal, we need to develop a logical and computational foundation for control theory. I use the name ``descriptive control theory" to emphasize the overarching theme of using logic to express, analyze, and solve problems in control theory. The program is to study control theory problems through their logical encodings, and use such encodings as a portal for bringing in suitable computational engines to solve the problems. Besides providing a formal foundation for existing methods in control theory, an important theme is to show the strength of logical decision procedures for generalizing these methods from dealing with polynomial-time solvable problems to NP-hard problems, and as a result, towards handling nonlinear and hybrid dynamical systems. 
 
Tarski's surprising result on the decidability of real arithmetic initiated the use of logic methods for solving problems in Euclidean spaces. Through quantifier elimination, the difficult tasks of geometric theorem proving become mechanizable, which significantly influenced the theory in practical fields such as robotic manipulation. The high computational complexity of the logical problems is still a bottleneck for delivering practical solutions. A fundamental problem of applying Taski's results to general control problems is the limitation of expressiveness of real arithmetic. Difficulty becomes apparent when we use first-order logic to reason about nontrivial dynamical systems: the first-order theory of real arithmetic with trigonometric functions is already highly undecidable. Consider the simple question of whether a one-dimensional continuous process, governed by some differential equation $\dot{x} = f(x,t)$, can start from $x=0$ and reach $x=1$ in finite time. A direct first-order encoding would require a formula such as the following:
\begin{eqnarray*}
\exists x_0 \exists t \exists x_t\; \bigg(x_0 = 0 \;\wedge\; x_t = x_0 + \int_{0}^t f(x(s),s)\mathrm{d}s\; \wedge\; x_t = 1\bigg).
\end{eqnarray*}
Given that solutions of differential equations are rarely polynomials, and most likely not analytically solvable, we can not directly reason with such logic formulas, and a logical approach seems hardly useful. 

We have developed a framework to bypass this core difficulty. Noting that the reasoning of continuous systems naturally involves numerical errors, we realize that
\begin{eqnarray*}
\exists x_0 \exists t \exists x_t\; \bigg(|x_0| \leq \delta_1 \;\wedge\; |x_t - (x_0 + \int_{0}^t f(x(s),s)\mathrm{d}s)| \leq \delta_2\; \wedge\; |x_t - 1|\leq \delta_3\bigg)
\end{eqnarray*}
is what really matters in practice, for some choices of numerical error bounds $\delta_1, \delta_2, \delta_3\in\mathbb{Q}^+$. Formalizing this observation, we have developed the theory of {\em delta-decisions} over the reals~\cite{DBLP:conf/lics/GaoAC12,DBLP:conf/cade/GaoAC12}. In this new theory we relax the standard decision problem to ask whether a sentence is true or its ``delta-strengthening" is false. Namely, we allow one-sided, delta-bounded numerical errors in logical decisions. With this change, the decision problem for logic formulas with arbitrary numerically computable functions, in the exact sense as developed in computable analysis, becomes decidable in bounded domains. The complexity of the delta-decision problems are also comparable to their discrete counterparts. These results stand in sharp contrast to the negative results for the standard decision problems, and provide the basis for developing a logical approach of control theory. We give a brief technical review of the theory of delta-decisions in Section~\ref{delta-decide}. 

Equipped with an expressive logic and decidability results, we can encode a wide range of control problems and study them from logical and computational perspectives. In particular, the following three directions form the main themes of descriptive control theory: 
\begin{itemize}
\item {\bf Descriptive Complexity.} The goal is to formalize the problems in control theory and to understand their computational complexity. We use the descriptive complexity approach: define a suitable logical language to express the control problems, such that their ``practical" computational complexity can be easily derived through the descriptions. Here, the measure of ``practical" complexity is defined through delta-decisions of the logic formulas, which we will give more details below. 
\item {\bf Logical Foundation.} The goal is to seek a logical foundation for existing theorems and methods in control theory. A natural approach is to follow the program of reverse mathematics to characterize control theory in suitable subsystems of second-order arithmetic, and for decidable theorems, decide their proof complexity. Such a foundation would also reveal computational content in these theorems. Moreover, the proofs should be formalized in an interactive theorem prover, which can be the basis for certifying practical control designs. 
\item {\bf Computational Engine.} The goal is to develop practical decision procedures for the logic formulas involved, which can serve as general algorithms for solving the control problems. As logical decision procedures usually target at hard problems (NP-hard and beyond), they may significantly extend existing methods in control theory, which mostly rely on polynomial-time algorithms. Moreover, the decision procedures should always produce proofs that can be validated through interactive theorem provers, and thus guarantee correctness-by-construction.
\end{itemize}
In what follows, I will first cover the background in Section~\ref{back}, and then discuss these three components in detail in Section~\ref{dc}, \ref{lf}, \ref{ce}. For each topic, I will outline two specific goals. The goals will combine theoretical investigations and practical implementations~\ref{summary}. 

\section{Background}\label{back}

\subsection{Delta-Decidability over the Reals}\label{delta-decide}

In this section, we review our theory of delta-decisions over the reals~\cite{DBLP:conf/lics/GaoAC12}. The theory allows us to consider first-order formulas over the reals with arbitrary {\em Type 2 computable functions}, a notion that has been well-developed in computable analysis~\cite{CAbook,Kobook}. We will introduce the notion of Type 2 computability first, and then give the main results about delta-decisions. 

Following computable analysis, we can encode any real number as an infinite sequence of rational numbers. For each real number $x$, a {\em name} of $x$ is any function $\phi: \N\rightarrow \mathbb{D}$ that {binary-converges} to $x$, namely, 
$$\forall n\in \N, |\phi(n) - x|\leq 2^{-n}.$$
We can then compute a real function $f:\mathbb{R}\rightarrow\mathbb{R}$ if there is an oracle Turing machine $M$ that, given the name of any argument $x\in\mathbb{R}$ of the function, computes the name of its value $f(x)$ up to an arbitrary digit in the following way:
\begin{definition}[Type 2 Computable Functions]
A real function $f: \R\rightarrow \R$
is {\em Type 2 computable}, if there is a function-oracle Turing machine $M$ such that for every $x\in \mathbb{R}$ and every name $\phi$ of $x$, given any $i\in\mathbb{N}$, the machine uses $\phi$ as an oracle, and $n$ as the input, and computes a rational number $M^{\phi}(i)\in\mathbb{Q}$, such that $|M^{\phi}(i)-f(x)|\leq 2^{-i}$. In other words, $M$ computes a function $\psi$ that binary-converges to $f(x)$ as its representation. 
\end{definition}
A most important property of computable functions is that they must have a computable modulus of continuity. Most common continuous real functions are Type 2 computable, including: polynomials with computable coefficients, exponential, trigonometric, square root, and logarithm functions, absolute value, and solution functions of Lipschitz-continuous ordinary differential equations. The complexity of Type 2 computable functions is also well studied. Intuitively, a real function $f:[0,1]\rightarrow\mathbb{R}$ is (uniformly) $\mathsf{P}$-computable ($\mathsf{PSPACE}$-computable), if it is computable by an oracle Turing machine $M_{f}$ that halts in polynomial-time (polynomial-space) for every $i\in \mathbb{N}$ and every $\vec x\in \dom(f)$. We denote this class of functions as $\mathsf{P_{C[0,1]}}$. The definitions of other classes such as $\mathsf{NP_{C[0,1]}}, \mathsf{PSPACE_{C[0,1]}}$ are similar. Omitting the formal details, we point out that most common real functions reside in $\mathsf{P_{C[0,1]}}$: absolute value, polynomials, binary $\max$ and $\min$, $\exp$, and $\sin$ are all in $\mathsf{P_{C[0,1]}}$. Moreover, it has been shown that solutions of Lipschitz-continuous differential equations are $\mathsf{PSPACE_{C[0,1]}}$-complete~\cite{DBLP:journals/cc/Kawamura10,Kobook}.

We can now consider the first-order language $\lrf^1$ over the real numbers, which allows the use of arbitrary Type 2 computable functions. This language is rich enough for expressing a wide range of continuous and hybrid systems and their properties. 
We write $\mathcal{F}$ to denote an arbitrary collection of symbols representing Type 2 computable functions over $\mathbb{R}^n$ for various $n$. We always assume that $\mathcal{F}$ contains at least the constant $0$, unary negation, addition, and the absolute value. (Constants are seen as constant functions.) Let $\mathcal{L_{\mathcal{F}}}$ be the signature $\langle \mathcal{F}, >\rangle$. $\mathcal{L}_{\mathcal{F}}$-formulas are always evaluated in the standard way over the corresponding structure $\mathbb{R}_{\mathcal{F}}= \langle \mathbb{R}, \mathcal{F}, >\rangle$.  It is not hard to see that we only need to use atomic formulas of the form $t(x_1,...,x_n)>0$ or $t(x_1,...,x_n)\geq 0,$ where $t(x_1,...,x_n)$ are built up from functions in $\mathcal{F}$. We can give an explicit definition of $\mathcal{L}_{\mathcal{F}}$-formulas as follows.
\begin{definition}[$\mathcal{L}_{\mathcal{F}}$-Formulas]
Let $\mathcal{F}$ be a collection of Type 2 functions, which contains at least $0$, unary negation -, addition $+$, and absolute value $|\cdot|$. We define:
\begin{eqnarray*}
&&t := x \; | \; f(t(x_1, ..., x_n)), \mbox{ where }f\in \mathcal{F}\mbox{, possibly constant};\\
&&\varphi := t> 0 \; | \; t\geq 0 \; | \; \varphi\wedge\varphi \; | \; \varphi\vee\varphi \; | \; \exists x\varphi \; |\; \forall x\varphi.
\end{eqnarray*}
We use the notation of {\em bounded quantifiers}, defined as $\exists^{[u,v]}x.\varphi =_{df}\exists x. ( u \leq x \land x \leq v \wedge \varphi)$ and $\forall^{[u,v]}x.\varphi =_{df} \forall x. ( (u \leq x \land x \leq v) \rightarrow \varphi)$. We say a sentence is bounded if it only involves bounded quantifiers. 
\end{definition}
On this normal form of $\lrf$-formulas, we can then define the $\delta$-variant of them as follows. 
\begin{definition}[$\delta$-Variants]
Let $\delta\in \mathbb{Q}^+\cup\{0\}$, and $\varphi$ a bounded $\mathcal{L}_{\mathcal{F}}$-sentence of the form
$$\varphi:\ Q_1^{I_1}x_1\cdots Q_n^{I_n}x_n.\psi[t_i(\vec x, \vec y)>0; t_j(\vec x, \vec y)\geq 0],$$
where $i\in\{1,...k\}$ and $j\in\{k+1,...,j\}$. The {\em $\delta$-strengthening} $\varphi^{+\delta}$ of $\varphi$ is defined to be the result of replacing each atomic formula $t_i > 0$ by $t_i > \delta$ and each atomic formula $t_j \geq 0$ by $t_j \geq \delta$, that is,
$$\varphi^{+\delta}:\ Q_1^{I_1}x_1\cdots Q_n^{I_n}x_n.\psi[t_i(\vec x, \vec y)>\delta; t_j(\vec x, \vec y)\geq \delta],$$
where $i\in\{1,...k\}$ and $j\in\{k+1,...,j\}$.
Similarly, the {\em $\delta$-weakening} $\varphi^{-\delta}$ of $\varphi$ is defined to be the result of replacing each atomic formula $t_i > 0$ by $t_i > -\delta$ and each atomic formula $t_j \geq 0$ by $t_j \geq -\delta$, that is,
$$\varphi^{-\delta}:\ Q_1^{I_1}x_1\cdots Q_n^{I_n}x_n.\psi[t_i(\vec x, \vec y)>-\delta; t_j(\vec x, \vec y)\geq -\delta].$$
\end{definition}
Note that in the definition, the bounds on the quantifiers are not changed. Our main theorem is the following. 
\begin{theorem}[$\delta$-Decidability]\label{main}
There is an algorithm which, given any bounded $\mathcal{L}_{\mathcal{F}}$-sentence $\varphi$ and $\delta\in \mathbb{Q}^+$, correctly returns one of the following two answers:
\begin{itemize}
\item ``$\mathsf{True}$'': $\varphi$ is true. 
\item ``$\delta$-$\mathsf{False}$": $\varphi^{+\delta}$ is false. 
\end{itemize}
Equivalently, there is an algorithm which, given any bounded $\varphi$ and $\delta\in \mathbb{Q}^+$, correctly returns one of the following two answers:
\begin{itemize}
\item ``$\delta$-$\mathsf{True}$'': $\varphi^{-\delta}$ is true. 
\item ``$\mathsf{False}$'': $\varphi$ is false. 
\end{itemize}
\end{theorem}
Note that the two cases can overlap. If $\varphi$ is true and $\varphi^{+\delta}$ is false, then the algorithm is allowed to return either one. The proof idea is that for any formula $\varphi$, the strictification of $\varphi$ is equivalent to the formula $\alpha(\varphi) > 0$. Whether this holds cannot, in general, be determined algorithmically, But given a small $\delta$, we {\em can} make a choice between the overlapping alternatives $\alpha(\varphi) > 0$ and $\alpha(\varphi) < \delta$, and this is enough to solve the relaxed decision problem. The dual versions of the theorem are useful in different contexts. The weakening is used for confirming that a certain property holds. The strengthening is used for synthesizing parameters. The differences will be discussed in later sections. We also have complexity results such as follows.
\begin{theorem}\label{compmain}
Let $\mathcal{F}$ be a class of computable functions. Let $S$ be a class of $\mathcal{L}_{\mathcal{F}}$-sentences, such that for any $\varphi$ in $S$, the terms in $\varphi_{[0,1]}$ are computable in complexity class $\mathsf{C}$ where $\mathsf{P_{C[0,1]}\subseteq \mathsf{C}\subseteq \mathsf{PSPACE_{C[0,1]}}}$. Then, for any $\delta\in \mathbb{Q}^+$, the $\delta$-decision problem for bounded $\Sigma_n$-sentences in $S$ is in $\mathsf{(\Sigma_n^P)^C}$.
\end{theorem}
As corollaries, we have the following completeness results for signatures of interest. 
\begin{corollary}
Let $\mathcal{F}$ be a set of $\mathsf{P}$-computable functions (which, for instance, includes $\exp$ and $\sin$). The $\delta$-decision problem bounded $\Sigma_n$-sentences in $\mathcal{L}_{\mathcal{F}}$ is $\mathsf{\Sigma_n^P}$-complete. 
\end{corollary}
\begin{corollary}
Suppose $\mathcal{F}$ consists of Lipschitz-continuous ODEs over compact domains. The $\delta$-decision problem for bounded $\mathcal{L}_{\mathcal{F}}$-sentences is $\mathsf{PSPACE}$-complete. 
\end{corollary}
Note that these complexity results stand in high contrast to the standard undecidability of the problems. In fact, they bring the hope of solving these formulas with practical decision procedures, as they are not beyond the complexity capacity of SAT and SMT solvers. Based on the theory, we have implemented a practical solver dReal that solves a wide range of nonlinear formulas, which will be discussed in Section~\ref{dreal}. 

\subsection{Control Theory} 

Control theory studies methods of regulating dynamical systems to achieve desired goals. Some detailed examples and results of control theory will accompany the discussion when needed. The main topics in control theory are: 
\begin{itemize}
\item Analyze the properties of a dynamical system under external controls, such as their stability, controllability, and observability. 
\item Design controllers that can regulate a dynamical system to satisfy certain specifications, such as stabilizing a system. 
\end{itemize}
A complete theory for linear dynamical systems has been well developed. The behavior of a linear system, such as stability and controllability, can be completely understood through properties of the matrix that determines its dynamics. However, beyond linear systems, existing methods have significant difficulty in dealing with nonlinear and hybrid systems. The goal of developing a logic-based methodology is to tackle the difficulty of existing methods. The logical approaches naturally bring in the power of computational engines that can extend existing methods from linear to nonlinear and hybrid systems. An important goal is to ensure that it is a strict extension: existing methods in control theory will be formalized and considered as algorithms for specific subclasses of the problems. The ultimate goal is to develop a new methodology that combines automation and certification. When the solving process is mechanized, checking the correctness of the mechanism becomes a simpler way of ensuring correctness of the results. 

\subsection{Computational Framework}\label{dreal}

The theory of delta-decidability not only provides a new perspective to look at decision problems over the reals, but also guides the development of concrete algorithms for solving the delta-decision problems. We say an algorithm is {\em delta-complete}, if it always terminates with correct delta-decisions. We have formally analyzed constraint solving algorithms and formulated formal conditions under which they are delta-complete. By combining such algorithms with logical decision procedures, we obtain decision procedures that can exploit the most out of both symbolic and numerical algorithms. 

Based on the theory, we have developed a practical solver dReal~\cite{DBLP:conf/cade/GaoKC13,DBLP:conf/fmcad/GaoKC13}~\footnote{\url{http://dreal.cs.cmu.edu}} that solves $\Sigma_1$-formulas in nonlinear theories over the reals in the framework of delta-complete decision procedures. Since 2012, dReal has been a very active project that combines the effort many collaborators~\footnote{\url{http://github.com/dreal}} and successfully applied to many practical problems~\cite{DBLP:conf/cade/GaoKC13,DBLP:conf/fmcad/GaoKC13,DBLP:conf/hybrid/KapinskiDSA14,DBLP:conf/fmcad/GaoKC13}. An important feature of the dReal tool is that it produces logical proofs to certify its answers. Such proofs can be used to produce certification of the correctness of concrete control designs, which can be easily checked in an interactive theorem prover. 

\section{Descriptive Complexity}\label{dc}

The first goal is to develop a logical description of problems in control theory. This step is the basis for both a logical foundation and a computational treatment of the problems. An immediate benefit of the encoding follows from the theory of delta-decisions over the reals: upper bounds on the complexity of these problems can be easily derived through their encoding in $\lrf^1$. For control design problems, goal is usually to find a function that satisfies certain goals. These problems are most naturally expressed with second-order quantification, which is the motivation for investigating second-order theories written within $\lrf^2$. 

I will first explain the approach of using first-order delta-decisions to encode control problems and deriving their complexity bounds, and then discuss the possibility of extending to a second-order formalism. 

\paragraph{Related Work.} Computational properties of most control problems are not clearly understood. The computational complexity of stability properties has been a topic of much recent investigation~\cite{DBLP:journals/corr/AhmadiP13,DBLP:journals/automatica/BlondelT99,DBLP:journals/automatica/BlondelT00,AAAthesis,DBLP:conf/hybrid/PrabhakarV13,DBLP:journals/corr/abs-1210-7420}. A focus of existing work is to establish various hardness results, i.e., lower bounds on complexity. It is shown that stability of simple systems is hard or impossible to solve algorithmically. Such results are proved by reducing combinatorial problems over graphs or matrices to stability problems, which can be analyzed with techniques of standard complexity theory. A limitation is that reduction techniques are usually not suitable for establishing upper bounds on complexity, and indeed most questions about upper bounds are open~\cite{AAAthesis}. 

\subsection{Descriptive Complexity in $\lrf^1$} 

The majority of control problems ask for real functions that satisfy certain properties. 

We need to justify the focus on the $\delta$-perturbed version of the problems. The ability of reasoning with both $\delta$-weakening and delta-strengthening gives us the ability of choosing the right perturbation to use. 
\begin{itemize}
\item For the problems that are concerned with a positive property, such as stability, controllability, observability, the perturbation on the negative side strengthens the problems, and are in fact the problems that we would like to solve, rather than the precise ones. 
\item For problems that are concerned with finding a witness such that some property holds, such as an optimal controller, it is more important to solve the perturbation on the positive side. The interpretation would be that the synthesized plan 
\end{itemize}

The study of stability properties of dynamical systems, for instance, can be fully described in $\lrf^1$. Here is an example of how this approach can be applied (more details in~\cite{DBLP:journals/corr/GaoKC14}). 
\begin{example}
Following standard definition, a system is stable i.s.L. if given any $\varepsilon$, there exists $\delta$ such that for any initial value $x_0$ that is within $\delta$ from the origin, the system stays in $\varepsilon$-distance from the origin. Naturally, the $\lrf$-representation of stability in the sense of Lyapunov is encoded in the following way:
\begin{quote}
\vspace{-.5cm}
\begin{definition}[{\sf L\_stable}]
We define the $\lrf^1$-formula {\sf L\_stable} to be:
\begin{eqnarray*}
& &\forall^{[0,\infty)} \varepsilon\exists^{[0,\varepsilon]} \delta \forall^{[0,\infty)} t\forall x_0\forall x_t .\; (||x_0||<\delta \wedge x_t = \int_0^t f(s)ds + x_0 )\rightarrow ||x_t||<\varepsilon.
\end{eqnarray*}
The {\em bounded form} of {\sf L\_stable} is defined by bounding the quantifiers as:\begin{eqnarray*}
& &\forall^{[0, e]} \varepsilon\exists^{[0,\varepsilon]} \delta \forall^{[0,T]} t\forall^X x_0\forall^X x_t. \;(||x_0||<\delta \wedge x_t = \int_0^t f(s)ds + x_0 )\rightarrow ||x_t||<\varepsilon, 
\end{eqnarray*}
where $e, T\in \mathbb{R}^+$ and $X$ is a compact set.
\end{definition}
\end{quote}
We can then define the $\delta$-stability problem using the $\lrf$-representation:  
\begin{quote}
\vspace{-.5cm}
\begin{definition}[$\delta$-Stability i.s.L.]\label{sl}
The $\delta$-stability problem i.s.L. asks for one of the following answers:
\begin{itemize}
\item {\sf stable}: The system is stable i.s.L. ({\sf L\_stable} is true). 
\item {\sf $\delta$-unstable}: Some $\delta$-perturbation of {\sf L\_stable} is false. 
\end{itemize}
\end{definition}
\end{quote}
We defined the {\em bounded} $\delta$-stability problem by replacing {\sf L\_stable} with the bounded form of {\sf L\_stable} in the definition. Now, using the complexity of the formulas, we have the following complexity results for the bounded version of Lyapunov stability. 
\begin{quote}
\vspace{-.5cm}
\begin{theorem}[Complexity]
Suppose all terms in the $\lrf$-representation of a system are in Type 2 complexity class $\mathsf{C}$.  Then the bounded $\delta$-stability problem i.s.L. resides in complexity class $\mathsf{(\Pi^P_3)^C}$. 
\end{theorem}
\end{quote}
This completes the example of obtaining complexity.\qed
\end{example}
Note that in the example, the new problems are defined through delta-perturbations of their logical encoding. Thus the new definitions are dependent on the descriptive approach. The same methodology can be extended to a wide range of topics in control theory. The fields of optimal, robust, adaptive control design, which are themselves fields in control theory, provides plenty of opportunity for logical formalization.

\subsection{Descriptions in $\lrf^2$}\label{second-order}

For many control design problems, their direct encoding requires a second-order language that naturally extends $\lrf^1$ with second-order quantifiers, which we call $\lrf^2$. For instance, consider the problem of finding an optimal controller for a system $\dot x = f({x}, u)$, with a cost function $g(x(t),u(t))$. Such a controller exists if the following $\lrf^2$-formula is true:
\begin{eqnarray*}
& &\hspace{-.7cm}\exists U \forall U' \forall {x_0}\forall {x_t} \forall {x_t'}\\
& &\hspace{-.7cm}\Bigg( \Big({x_t} = {x_0} + \int_{0}^t f({x}(s),U(s))\mathrm{d}s \wedge\;  x_t' = x_0 + \int_{0}^t f(x(s),U(s))\mathrm{d}s
\;\wedge\; \phi(x_0, x_t) \;\wedge\; \phi(x_0, x_t')\Big)\\
& &\hspace{5.7cm}\rightarrow \Big(\int_0^t g(x(s),U(s))\mathrm{d}s \leq  \int_0^t g(x(s),U'(s))\mathrm{d}s)\Big) \Bigg).
\end{eqnarray*}
$U$ and $U'$ denote control functions, and $\phi$ encodes some constraints on the initial and end states. The formula states that there exists a control function $U$ such that any other control function $U'$ that achieves the same goals would cost more than $U$, with respect to the cost function $g$. 

We have conjectured in~\cite{DBLP:conf/lics/GaoAC12} that the second-order language $\lrf^2$ is still delta-decidable under suitable interpretations of the second-order variables, since techniques for proving $\delta$-decidability in the first-order case should apply to arbitrary compact metric spaces. We need to develop complexity analysis for $\lrf^2$, which should systematically extend existing complexity results in computable analysis from real functions to functionals. 

\subsection{Main Goals in Descriptive Complexity} 
In sum, the two main goals under the theme of descriptive complexity are the following:
\begin{goal}
Express control theory in $\mathcal{L}^2_{\mathbb{R}_{\mathcal{F}}}$ with a suitable interpretation (by choosing appropriate domains for the second-order variables), whose bounded $\delta$-decision problem should be decidable.
\end{goal}
\begin{goal}
Develop a descriptive complexity theory for $\lrf^2$-formulas with respect to the $\delta$-decisions and prove decidability results, possibility with preliminary complexity results. 
\end{goal}
The goals are important for the investigations in the other two components of the main theme. With the logical encodings, we can investigate the strength of theorems in control theory regarding these properties. At the same time, the encodings place the problems into different computational hierarchies, for which we can apply practical computational engines to solve. 

\section{Logical Foundation}\label{lf}

The next task is to develop a logical foundation for main results in control theory. Mathematically, a formal foundation of control methods provides a basis for developing rigorous methodology. The goal is to develop a formal basis of control theory in suitable subsystems of second-order arithmetic, following the program of reverse mathematics. The benefit of focusing on weak theories, compared to ZFC or higher, is that we can analyze the proofs such that their computational content becomes clear. Computationally, formal proofs of main results in control theory can be used to certify correctness of  concrete system designs. 

\paragraph{Related Work. } There are various approaches in providing logic-based approaches to dynamical systems and control theory. Most of the existing work focuses on expressive nonclassical languages that can encode complex behaviors of dynamical systems~\cite{DBLP:conf/lics/Platzer12}. Our focus is to develop a formal foundation for the proofs of the main theorems, which has not been done before. Most of core of control theory only requires a classical first-order to second-order language. Once the study of the core problems is clear, more language constructs, such as temporal modalities, can be further introduced. 

\subsection{Proof Complexity and Reverse Mathematics} 

It is an interesting question to know the mathematical commitment that we have when proving facts about dynamical systems. Theorems in control theory typically give conditions about when a system satisfies certain control properties. There are two sides to control theory. One is the analysis side, and the other is the algebraic side. The proof complexity is lower on the algebraic side, which imposes stronger assumptions on the structure of the systems, for instance, linear systems represented as matrices. The analysis side requires stronger theories. 

One presumably easy but still interesting first task is to study the reverse mathematics of the state-space control theory for linear dynamical systems. It is mostly clear that $\mathsf{RCA}_0$ is enough to develop much of linear algebra~\cite{Simpson}. Indeed, following the work in proof complexity~\cite{cookbook}, restricted forms of many facts from linear algebra, such as the Cayley-Hamilton theorem, have polynomial-time proofs. Most of the control theory for linear dynamical systems are stated in the language of matrices. For instance, consider the {\em Kalman test for controllability}. The Kalman test for controllability states that an $n$-dimensional linear system $\dot {\bf x} = A{\bf x}+B{\bf u}$ is {\em controllable} iff the {\em Kalman matrix}
$[B\ AB\ \cdots\ A^{n-1}B]$ is of rank $n$. Such theorems are quite straightforward consequences of the Cayley-Hamilton theorem. Their proof complexity should not be high. 

The analysis aspect of control theory studies dynamical systems through properties of solutions of differential equations. For instance, consider the Lyapunov method for nonlinear stability as follows. 
\begin{example}
Recall that the definition of stability in the sense of Lyapunov is given by the following $\lrf^1$-formula:
\begin{eqnarray*}
\varphi_S:\ \forall\varepsilon\exists^{[0,\varepsilon]} \delta \forall^{[0,\infty)} t\forall x_0\forall x_t .\; (||x_0||<\delta \wedge x_t = \int_0^t f(s)ds + x_0 )\rightarrow ||x_t||<\varepsilon.
\end{eqnarray*}
Let $V(p,x)$ be a function, parameterized by $p\in \mathbb{R}$, whose partial derivative ${\partial V}/{\partial x}$ is a Type 2 computable function. Let $D$ be the parameter space for $p$ and $X$ be the state space of $x$. We then have the following $\lrf$-formula is a sufficient condition for stability in the sense of Lyapunov
$$\varphi_L:\ \exists p\forall x\; \bigg(V(p,x)\geq 0 \wedge V(p,0) = 0\wedge \frac{\partial V(p,x)}{\partial x}f(x)\leq 0\bigg)$$
The Lyapunov methods is based on the proof that $\varphi_L\rightarrow \varphi_S$ holds as a theorem over the reals, which can be analyzed in a suitable subsystem of second-order arithmetic. \qed
\end{example}
Analysis methods cover most of the content in nonlinear control. The classical aspects of control theory based on frequency-domain analysis is worth investigating as well, which involves more complex analysis, and functional analysis. Many valuable results in this direction have been obtained in the work of Yokoyama~\cite{yoko}. For instance, Yokoyama showed that uniformly convergence of Fourier series for $C^1$-functions and $L^2$-convergence of Fourier series for continuous functions are equivalent to $\mathsf{WKL}_0$ over $\mathsf{RCA}_0$. In this perspective, it is interesting to understand whether modern control theory, while being more powerful, depends on weaker axioms than classical control theory. If that is the case, this would be a clear sign of improvement in the development of control theory. 

\subsection{Formal Proofs}

An important part of establishing a logical foundation for control theory is the formalization of the proofs in an interactive theorem prover. The interactive proofs do not necessarily start from the weak subsystems, but the formalization in the previous section will certainly be valuable for the formalization of the proofs. 

Candidate proof assistants include Coq, HOL, Isabel. Both the algebraic side and analysis side of control theory can find basic packages to start with. For instance, as a core theorem in matrix algebra, the Cayley-Hamilton theorem is proved in Coq and HOL. It is the basis of deriving other parts of the theory of linear systems. 

Different consideration seems to favor different systems for developing the formalization. HOL has concise formulation, while Coq seems to be attractive to a wider community and would be a good choice for the future steps of the integration into proofs of correctness of control software. Another option is to work on top of the new theorem prover ``Lean" (\url{http://leanprover.net}), a project recently started by Leonardo de Moura. The design principle is to have a framework that can use automatic solvers at the backend as engines in the proofs as much as possible. The benefit of working with Lean is that I can take an active role in developing the theorem prover itself which can help the development of a control theory library. 

With a suitable choice of the proof environment, the goal is to have a control theory library. An ideal plan is to produce a textbook-like document that develops the core part of control theory with all theorems and proofs formalized in the proof assistant of choice. Theorems in control theory that requires formalization includes stability, observability, controllability for linear systems as previously mentioned, Lyapunov theory for nonlinear systems, and various results in optimal, robust, and adaptive control. 

Besides the mathematical value of such formalization, An important use of the formal proofs is computational: these proofs can serve as a basis for formal verification of practical control designs. An interesting question is how to connect proofs of the main theorems to automated proofs of the concrete designs. 

\subsection{Main Goals in Logical Foundation}
\begin{goal}
Categorize theorems in control theory in suitable subsystems of second-order arithmetic, or find proof complexity for decidable theorems concerning linear systems. 
\end{goal}
\begin{goal}
Formalize the main theorems in control theory in an interactive theorem prover.  
\end{goal}
\section{Computational Engine}\label{ce}

The main benefit of using a logical approach to express and formalize control theory is that it is backed by the use of computational engines. The methodology becomes that we can express a problem formally, and then the problem is solved using decision procedures for the logic theory. In this way, decision procedures for the logic formulas become generic algorithms for the control problems. 

It is important to make sure that existing methods in control theory can be used as partial algorithms for subclasses of these formulas. A challenge is to ensure that algorithms based on matrix operations and convex optimization, for instance, can all be suitably called by the logic solver. To make sure of this in a solver, this requires detailed analysis of the numerical algorithms so that delta-completeness can be ensured. The additional benefit, besides automation, is that these decision procedures should automatically produce witnesses or proofs for their answers, such that the correctness of the full control design can be certified with a formal proof. For sat answers, one can simply plug in the solutions and check the correctness (up to delta-bounded errors). For unsat answers, one has to produce a proof of refutations that have been found by the solver. This requires detailed analysis of numerical algorithms.

In the next steps, a main target is $\Sigma_2$-sentences in $\lrf^1$, and also to handle important classes in $\lrf^2$, at least up to the point that the existing algorithms in control design can be used. This involves the use of methods from calculus of variations, dynamic programming, etc. An interesting point to note is that the problems here all become generalizations of optimization problems. Scalar optimization problems corresponds to $\exists\forall$ formulas with one single existentially quantified variable. 

\subsection{From $\Sigma_1$ to $\Sigma_2$ Problems} 

As mentioned in the background section, we have developed the framework of {\em $\delta$-decision procedures} for solving $\Sigma_1$-sentences in $\lrf^1$~\cite{DBLP:conf/cade/GaoAC12}, and implemented a practical solver dReal based on the theory. Many standard control-theoretic problems, such as the validation of Lyapunov functions for nonlinear systems can be straightforwardly encoded as $\Sigma_1$-sentences, for which dReal has been used in practical problems~\cite{DBLP:conf/hybrid/KapinskiDSA14}.

The important next step of research is to solve $\Sigma_2$-sentences, which will be able to encode a wide range of general control problems. Consider Lyapunov analysis of systems as an example. 
\begin{example}
The search of Lyapunov functions can be done by fixing a template of Lyapunov functions and search for parameters. 
Let $D$ be the parameter space for $p$ and $X$ be the state space of $x$. We then have the following $\lrf$-formula is a sufficient condition for stability in the sense of Lyapunov
$$\varphi_L:\ \exists p\forall x\; \bigg(V(p,x)\geq 0 \wedge V(p,0) = 0\wedge \frac{\partial V(p,x)}{\partial x}f(x)\leq 0\bigg)$$
If $p$ is a single parameter, the problem can be solved by binary search on $p$. In general, if $p$ is a vector, we need procedures for solving $\Sigma_2$ problems. \qed
\end{example}
General $\Sigma_2$ problems can be seen as a generalization of vector optimization problems.  Decision procedures for such sentences can be developed through recursive calls to the algorithms for the existential sentences, and also exploit existing optimization algorithms:
\begin{itemize}
\item The first one is to use optimization solvers as oracles, and solve problems in the DPLL(T) way. This direction should extend all benefits of existing optimization algorithms. Ideally, this provides a framework that strictly generalizes existing practice in control theory. The key is to formalize the numerical procedures such that delta-completeness is achieved. 
\item The second one is the generic procedure of calling the solver itself recursively. One would need to first leave the existentially quantified variables as free, and solve the universally quantified constraints first. The results are then used for pruning the constraints on the existentially quantified variables. 
\end{itemize}
A challenge is that when we solve for the innermost universally quantified constraints, we need over-approximation on the values of the negation of the formulas. This requires algorithms for computing under approximations for the real variables. Theoretically, this is not harder than solving based on over approximation, but practical algorithms are yet to be developed. On the other hand, we can formulate a notion of mixing delta strengthening and weakening for applications.

\subsection{Towards $\lrf^2$} 

The next step is to develop algorithms for solving second-order formulas in $\lrf^2$. As discussed in Section~\ref{second-order}, control design problems can usually be encoded as the search for functions that satisfy certain constraints. For instance, optimal control problems search for functions that minimize a functional that represents cost. Obviously, it is computationally difficult to have a generic algorithm. There is no existing work on computing second-order formulas over the reals. Still, there are three directions to approach the problem. 

One straightforward approach is to restrict the search with templates, which amounts to restricting the interpretation of the second-order variables, and reduce the problems to a series of first-order $\Sigma_2$ problems. This is in fact the commonly used strategy in control design. For instance, when designing a PID (Proportional-Integral-Derivative) controller, we would use the template 
$$\mathrm{u}(t)= K_p{e(t)} + K_{i}\int_{0}^{t}{e(\tau)}\,{d\tau} + K_{d}\frac{d}{dt}e(t)$$
and search for the parameters $K_p, K_i, K_d\in \mathbb{R}$. Another example is in Lyapunov analysis, in which one typically search through polynomials of increasingly higher degree. 

The second approach is to work with parametric equations of a function, which can be considered as a curve, and numerically compute the values of the parameters to traverse on the curve. This is indeed what we have done for formulas with ODEs, when dealing with reachability problems for hybrid dynamical systems~\cite{DBLP:conf/fmcad/GaoKC13}. In this case, we compute the solution of the ODE is parameterized by the time variable, and produce a complete trace. The piecewise linear trace is a delta-approximation of the exact function. We need to formalize this approach and further investigate its usability and limits. 

The third direction is to incorporate methods from infinite-dimensional optimization, such as calculus of variations and dynamic programming. The algorithms typically involve solving partial differential equations for solving such problems. Formalization of the existing algorithms would require a significant amount of work, which will be an important step towards the merging of symbolic and numerical algorithms. 

\subsection{Main Goals in Computational Engine}
\begin{goal}
Develop practical algorithms for the $\delta$-decision problem of $\exists\forall$-sentences in $\lrf^1$.
\end{goal}
\begin{goal}
Develop a framework for solving $\delta$-decision problems of $\lrf^2$-sentences.
\end{goal}
\section{Summary}\label{summary}

The six goals in the previous sections can be divided into theoretical and computational goals. The theoretical ones include:
\begin{enumerate}
\item Completely formalize first-order problems in control theory in $\lrf^1$, and characterize the complexity accordingly. 
\item Develop a suitable second-order language $\lrf^2$ and prove decidability for control design problems that can be expressed accordingly. 
\item Categorize theorems in control theory in suitable subsystems of second-order arithmetic, and find proof complexity for decidable theorems. 
\end{enumerate}
And the computational ones are:
\begin{enumerate}
\item Develop formal proofs of basic theorems in control theory in an interactive theorem prove.
\item Implement a practical solver for $\Sigma_2$ problems in control theory. 
\item Develop a framework for solving second-order problems. 
\end{enumerate}
\bibliographystyle{abbrv}
\bibliography{ref}

\end{document}